\documentstyle[12pt]{article}

\advance\textwidth 2cm
\advance\oddsidemargin -1cm
\def\Kuchar{Kucha\v r}
\def\Hajicek{H\' aj\'\i\v cek}
\def\Schrodinger{Schr\" odinger}
\newcommand\beq{\begin{equation}}
\newcommand\eeq{\end{equation}}

\def\obar{\overline}

\def\la{{\lambda}}

\begin{document}
\baselineskip=24pt
\title{The Global Problem of Time}
\author{Arlen Anderson\thanks{arley@physics.mcgill.ca}\\
Department of Physics\\
McGill University\\
Montr\'eal PQ H3A 2T8 Canada}
\date{March 27, 1992}
\maketitle
\vspace{-10cm}
\hfill McGill 92-15

\hfill hep-th/9205112
\vspace{10cm}
\begin{abstract}
Time does not obviously appear amongst the coordinates on the constrained
phase space of general relativity in the Hamiltonian formulation. Recent
work in finite-dimensional models claims that topological obstructions
generically make the global definition of time impossible. It is shown
here that a time coordinate can be globally defined on a constrained
phase space by patching together local time coordinates, just as coordinates
are defined on topologically non-trivial manifolds.
\end{abstract}
\newpage

In the Hamiltonian formulation of general relativity, instead of dynamical
equations, one finds four initial value constraints, the super-Hamiltonian
and super-momentum constraints. The coordinates on the phase space consist
of the 3-metric and the extrinsic curvature of a spacelike hypersurface,
and the constraints restrict evolution to a subset of phase space. The
obvious question is: where is time? Is there a coordinate on the
constrained phase space which can be identified as time? Recent work claims
that there are topological obstructions which make such an identification
impossible\cite{Haj,Tor}. We shall argue here that a global definition of
time can be found by using canonical transformations to define local
time coordinates and then to patch them together to cover the constrained
phase space, much as
coordinates are defined on topologically non-trivial manifolds.

Few would doubt the existence of time, but as a question about the
constrained phase space of general relativity, the answer is less
self-evident. The problem is not academic.
While the global definition of time on the
constrained phase space is a purely classical question, the whole program
of canonical quantization depends on its successful resolution. If time
could not be defined globally, how could one speak about evolution, let
alone define a Hilbert space with a conserved inner product?

One might have hoped that a \Schrodinger-type super-Hamiltonian
\beq
{\cal H}=p_t+H(p_k,q_k,t)=0.
\eeq
would have followed from the Hamiltonian formulation of general relativity.
Time then would have been easily identified as the variable conjugate to
$p_t$, and quantization would be straightforward. This not being the case,
the natural question is: is there a canonical transformation from the
Wheeler-DeWitt super-Hamiltonian to a \Schrodinger-type super-Hamiltonian
such that the two are fully equivalent? This is the global problem of time,
as formulated by \Kuchar\cite{Kuc}.

The infinite-dimensional nature of general relativity makes it as yet
too difficult to handle directly, but considerable insight can be
obtained from finite-dimensional models.
There, the answer to \Kuchar's question is simply: no, in general there is no
canonical transformation from an arbitrary super-Hamiltonian to one of
\Schrodinger-type such that
the resulting system is {\it fully} equivalent to the original one. This
result is not however surprising and, in fact, to insist by it that the
global problem of time is insoluble misses the spirit of the question.
Arguably the distinguishing characteristic of time is that it is a
coordinate whose each and every value occurs exactly once along each of all
possible classical trajectories. This property is familiar in
\Schrodinger-type systems. The global problem of time may be restated: is
there a coordinate on the constrained phase space of the Wheeler-DeWitt
equation which, because it increases monotonically along every classical
trajectory, can serve as time?

In the finite-dimensional case, the answer is now yes, but the time
coordinate must be defined locally and
then patched together to cover the full constrained phase space. The
situation is completely analogous to defining coordinates on a manifold.
One asks if there is a map between the manifold and Euclidean space.
Generically there is no map such that Euclidean space is fully equivalent
(isomorphic) to the original manifold--consider, for example, compact
manifolds. Nevertheless, one can define coordinates globally by finding
maps locally between the manifold and Euclidean space and patching these
maps together by giving transformations between them in regions where they
overlap. The global problem of time concerns defining a coordinate on the
constrained phase space of the super-Hamiltonian which can be identified as
time. It should come as no surprise that the non-trivial topology of this
constraint surface may require that the time coordinate be defined on
patches.

\Hajicek\cite{Haj} has extensively studied the global problem of time in
the finite-dimensional context. He formulates the problem in terms of the
existence of a hypersurface such that every classical trajectory intersects
it exactly once. This corresponds to identifying a common unique instant of
time amongst all classical trajectories. If no such hypersurface can be
found, \Hajicek\ argues that time cannot be globally defined.

For most super-Hamiltonian constraints on a finite-dimensional phase space,
\Hajicek\ finds that there are topological obstructions to the existence of
such a hypersurface and thus to a globally defined time. These obstructions
take three forms. First, the super-Hamiltonian may have fixed points. These
are classical trajectories which consist of a single point in phase space.
As such, successive instants of time are indistinguishable and all lie on
the same hypersurface. Second, the super-Hamiltonian may have (almost)
periodic trajectories. In this case, trajectories repeatedly intersect a
candidate hypersurface. Finally, there may be non-Hausdorff pairs of
trajectories. These are trajectories who always share a common neighboring
trajectory, but are not themselves neighbors, so that if a hypersurface
were to intersect them both once, it would have to intersect a neighboring
trajectory more than once. Typically this happens at unstable fixed
points.

A \Schrodinger-type super-Hamiltonian has none of these obstructions.
It is immediate then that if a given super-Hamiltonian has any of them,
it cannot
be fully equivalent to a \Schrodinger-type Hamiltonian. But, to repeat,
this does not mean that time cannot be globally defined.
Furthermore, these obstructions are not pathological.  Flat space itself is
a fixed point of the Wheeler-DeWitt equation, the super-Hamiltonian of
general relativity.

In order to understand how time is globally defined in the presence
of topological obstructions,
it is instructive to consider a super-Hamiltonian constraint proposed
by \Hajicek
\beq
\label{sH}
{\cal H}={\textstyle{1\over 2}}(-p_0^2 +q_0^2 +p_1^2 -q_1^2)=0.
\eeq
This super-Hamiltonian has an unstable fixed point at $p_0=0=q_0=p_1=q_1$,
and several pairs of non-Hausdorff orbits, for example, the pair of
trajectories ($\la$ is the affine parameter labelling evolution in the
constrained phase space)
\begin{eqnarray}
p_1=0=q_1,& p_0=e^{\la},\ \ q_0=-e^{\la} \\
p_1=0=q_1,& p_0=e^{-\la},\ \ q_0=e^{-\la} . \nonumber
\end{eqnarray}

Since the super-Hamiltonian is separable, it can clearly be reduced to
action-angle variables by canonical transformation.  Choosing to do so
only with the $(p_0,q_0)$ variables, the canonical transformation
\begin{eqnarray}
\label{c1}
\obar p_0&=&{\textstyle{1\over 2}}(-p_0^2+q_0^2), \\
\obar q_0&=&-\ln (p_0+q_0), \nonumber
\end{eqnarray}
reduces the super-Hamiltonian to \Schrodinger\ form
\beq
{\cal H}_s=\obar p_0 +{\textstyle{1\over 2}}(p_1^2 -q_1^2)=0.
\eeq

The \Schrodinger\ super-Hamiltonian has no fixed points and no non-Hausdorff
pairs of trajectories, so it is not fully equivalent to the original
super-Hamiltonian. The difficulties of the original super-Hamiltonian have
been transformed away by the canonical transformation: the fixed point has
been sent to infinity, along with one half of the pair of non-Hausdorff
trajectories. All of the remaining trajectories share a common time as
$\obar q_0$ is correlated linearly with the affine parameter $\la$.
Moreover, all super-Hamiltonians which can be reached by canonical
transformations which
preserve the linear form of $\obar p_0$ share this property, and none have
fixed points or non-Hausdorff pairs of orbits.

A second canonical transformation can be made which reaches the
trajectory missed by the first one.  This is the transformation
\begin{eqnarray}
\label{c2}
\obar p_0'&=&{\textstyle{1\over 2}}(-p_0^2+q_0^2), \\
\obar q_0'&=&\ln (-p_0+q_0), \nonumber
\end{eqnarray}
which again reduces the super-Hamiltonian to \Schrodinger\ form
\beq
{\cal H}_s'=\obar p_0' +{\textstyle{1\over 2}}(p_1^2 -q_1^2).
\eeq
This sends the other half of the non-Hausdorff pair to infinity while
bringing back the first.

The essential feature now is that these
two canonical transformations overlap for most
trajectories, and they give a canonical transformation between the two
time coordinates
\beq
\obar q_0'=\obar q_0 +\ln 2\obar p_0.
\eeq
Thus, time is defined globally by patching together these two locally
defined coordinates with this transition function.  Since the fixed point
does not change with time, both coordinates are valid to describe its
evolution.

Having seen how fixed points and non-Hausdorff pairs of trajectories are
affected by canonical transformation, there remains only the obstruction of
(almost) periodic trajectories.
Changing the sign of $q_0^2$ and $q_1^2$ in the
super-Hamiltonian (\ref{sH}) introduces periodic trajectories.  The
canonical transformations (\ref{c1}) and (\ref{c2}) with $q_0$ replaced
by $iq_0$ again give the reduction to the \Schrodinger\ form.  It is clear
what has happened.  The periodic trajectories are ``unwound'' by the
canonical transformation because of the many-sheeted nature of the
logarithm.  Effectively, a transformation has been made to a covering
space of the original space.  Time is again defined globally by the two
local coordinates $\obar q_0$ and $\obar q_0'$.

In this simple example, we have seen how the topological obstructions of
Hajicek are handled by canonical transformations.  These obstructions
deny the full equivalence of a generic super-Hamiltonian and one of
\Schrodinger-type, but this does not mean that time cannot be globally
defined on the constrained phase space of the super-Hamiltonian.  Rather
we find that time may be defined locally by canonical transformation to a
\Schrodinger-type super-Hamiltonian and the different local times patched
together by canonical transformations in regions where they overlap, much
as coordinates are defined on a topologically non-trivial manifold.  The
nature of the obstructions in a finite-dimensional phase space are such
that this can always be done.  Every indication is that this extends to
the infinite-dimensional case, resolving the global problem of time.

\vskip 0.5cm
Acknowledgements:  I would like to thank P. \Hajicek, K. \Kuchar, and R.
Myers for stimulating discussions.  This work was supported in part by
grants from the Natural Sciences and Engineering Research Council and
les Fonds FCAR
du Qu\'ebec.


\begin{thebibliography}{9}
\bibitem{Haj} P.\ Hajicek, Phys.\ Rev.\ D{\bf 34}, 1040 (1986); J. Math.
Phys. {\bf 30}, 2488 (1989); Class.\ Quantum Grav.\ {\bf 7}, 871 (1990);
M.\ Sch\"on and P.\ Hajicek, Class.\ Quantum Grav.\ {\bf 7}, 861 (1990).

\bibitem{Tor} C. Torre, Utah State Univ. preprint
FTG-110-USU/hep-th-9204014 (1992).

\bibitem{Kuc} K.\ \Kuchar, ``Time and Interpretations of Quantum Gravity,''
{\it Proceedings of the 4th Canadian Conference on General
Relativity and Relativistic Astrophysics}, eds. G. Kunstatter, D. Vincent
and J. Williams (World Scientific, Singapore, 1992).

\end{thebibliography}
\end{document}